# User-based key frame detection in social web video

**Konstantinos Chorianopoulos**
Ionian University, Greece
choko@ionio.gr

**ABSTRACT**
Video search results and suggested videos on web sites are represented with a video thumbnail, which is manually selected by the video up-loader among three randomly generated ones (e.g., YouTube). In contrast, we present a grounded user-based approach for automatically detecting interesting key-frames within a video through aggregated users' replay interactions with the video player. Previous research has focused on content-based systems that have the benefit of analyzing a video without user interactions, but they are monolithic, because the resulting video thumbnails are the same regardless of the user preferences. We constructed a user interest function, which is based on aggregate video replays, and analyzed hundreds of user interactions. We found that the local maximum of the replaying activity stands for the semantics of information rich videos, such as lecture, and how-to. The concept of user-based key-frame detection could be applied to any video on the web, in order to generate a user-based and dynamic video thumbnail in search results.

**ACM Classification Keywords**
H.5.1 Multimedia Information Systems

**INTRODUCTION**
Web search engines use video thumbnails to represent videos in search results. Moreover, web video sites provide video thumbnails to facilitate user's navigation between related videos. Nevertheless, most of the existing content-based techniques that extract thumbnails at regular time intervals, or from each shot/scene are inefficient, because there might be too many shots in a video (e.g., how-to video), or rather few (e.g., lecture video). For example, search results and suggested links in YouTube are represented with a thumbnail that the video authors have manually selected out of the three fixed ones (Figure 1). By analogy to the early web-text search engines that were based on author declaration of important keywords, the current video search engine approach puts too much trust on the video thumbnails selected by the video author. Besides the threat of authors tricking the system, the author-based approach does not consider the variability of users' knowledge and preferences. Thus, there is a need for selecting video thumbnails according to the collective action of video viewers, in order to represent a video with important video segments.

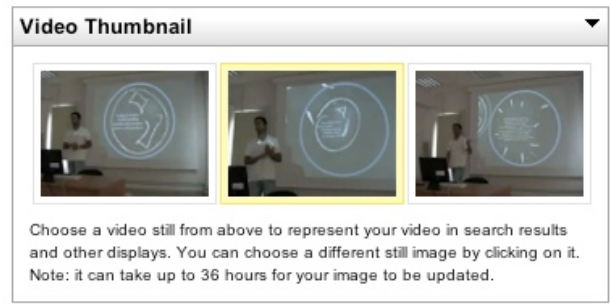

**Figure 1** The YouTube upload tool asks the user to manually select a video frame, which has been randomly generated.

Previous user-based research on web video has focused on the meaning of the comments, tags, re-mixes, and micro-blogs, but has not examined simple user interactions with a web-based video player. In the seminal user-based approach to web video, Shaw and Davis [8] proposed that video representation might be better modeled after the actual use made by the users. In this way, they have employed analysis of the annotations to understand media semantics. Peng et al. [6] have examined the physiological behavior (eye and head movement) of video users, in order to identify interesting key-frames, but this approach is not practical because it assumes that a video camera should be available and turned-on in the home environment. Shamma et al. [7] have created summaries of broadcasts (sports and political debate respectively) by analyzing the twitter stream of the respective real-time event. Although the twitter stream is very rich in meaning, it lacks the real-time accuracy that is required in the generation of video thumbnails, because it might take an arbitrary amount of time to type and send a text message. In contrast, the proposed method is only based on real-time user interactions, such as replay.

**METHODOLOGY**
The evaluation methodology consists of: 1) customized web video player that logs user interactions, 2) manually selected video segments of interest, 3) controlled experiment that produces a user interaction data-set, 4) heuristic for automatically generating video thumbnail.

The experimental web video player (Figure 2, left part) employs few buttons. There is the familiar pause/play button, but instead of the common video seek bar timeline, we employed two fixed-seek buttons. The GoBackward goes backward 30 seconds and its main purpose is to replay interesting parts of the video, while the Goforward button jumps forward 30 seconds and its main purpose is to skip insignificant video segments. Next to the player's button there is the current cue-time and the total time of the video in seconds. We did not use a random seek timeline because it would be difficult to analyze users' interactions. Moreover, Li et al. [4] observed that when seek thumb is used heavily, users had to make many attempts to find the desirable section of the video and thus caused significant delays.

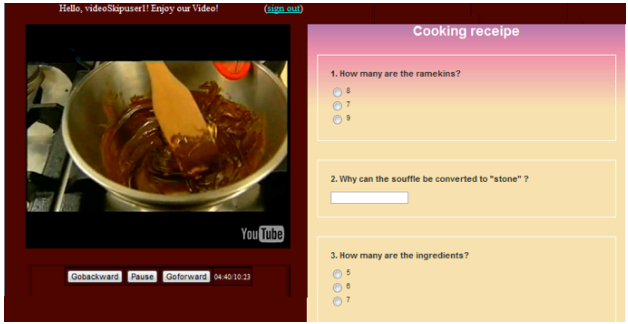

**Figure 2 The experimental web video player includes skipping buttons and questionnaire functionality.**

Instead of mining real video usage data, we have devised a controlled experiment, because it provides a clean set of data that might be easier to model and understand. We focused on videos that are as much visually unstructured as possible (e.g., lecture, how-to), because content-based algorithms have already been successful with those videos that have visually structured scene changes (e.g., movies, series). In order to experimentally replicate user interest, we added an electronic questionnaire (Figure 2, right part) that corresponds to a few manually selected video segments (semantics). According to Yu et al. [9] there are segments of a video clip that are commonly interesting to most users, and users might browse the respective parts of the video clip in searching for answers to some interesting questions.

In general, the baselines and controls for thumbnail selection and summarization are well known (both in their robustness and fragility) however, for this experiment, we assert such prior research is not a suitable control for this research study. That is to say, a human-centered approach identifies salient signals from human behavior, and not signals present in the video content [3], or its production itself [2]. More so, for the lecture and how-to videos we focus on, the video content tends to be either very static (usually a speaker at a podium), or very dynamic (multiple moving cameras provide alternative views of the same object and/or activity). Thus, signal processing and content analysis approaches tend to fail, because they produce few or too many key-frames respectively.

We chose to work with lecture and how-to videos and we selected a reference set of semantics, which we represented with the respective questions. The questions were relatively simple to answer, and did not depend on any previous knowledge, besides the information available within the video itself (Table 1). Therefore, the users had to seek/scrub through the video in order to answer those questions. It is expected that in a future field study, when enough user data is available, user behavior will exhibit similar patterns even if they are not explicitly asked to answer questions. This assumption might be especially valid in the case of informational videos (e.g., lectures, how-to), when users seek to find important information.

Our main interest is with lecture videos for two reasons: 1) they lack any meaningful visual structure that might have been helpful in the case of a content-based system, and 2) they contain lots of audio-visual (verbal and non-verbal) information that a user might actively seek to retrieve. In addition to video lecture, we employed a how-to (cooking) video because it has a rather complicated and active visual structure, which might have created too many false positives for a content-based approach.

| Video | Indicative questions |
|---|---|
| Lecture A | • Which are the main research topics?<br>• What the students did not like?<br>• What time does the first part of the talk end? |
| How-to B | • How many are the ramekins?<br>• How many are the ingredients?<br>• Which is the right order for mixing the ingredients? |

**Table 1 Example questions from each video. The questions are not supposed to be meaningful, but to direct the users towards a video segment.**

The experiment took place in a lab with Internet connection, general-purpose computers, and headphones. Twenty-three university students (18-35 years old, 13 women and 10 men) spent approximately ten minutes to watch each video (buttons were muted). All students had been attending the Human-Computer Interaction courses at the Department of Informatics (…) at a post- or under-graduate level and received course credit in the respective courses. Next, there was a time restriction of five minutes, in order to motivate the users to actively browse through the video and answer the respective questions. We did not directly encourage the users to actively seek, but we informed the users that the purpose of the study was to measure their performance in finding the answers to the questions within time constraints.

It is our main aim to examine whether the user interest function and the semantics are similar. In order to evaluate the performance, we modeled the user interest as a time series and we compared the observed user interest to the manually selected semantics that contained the answer to the question the users' were seeking for. Firstly, we considered that every video is associated with an array of k cells, where k is the duration of the video in seconds. Next, we modified the value of each cell by one, depending on the type of interaction. For each GoBackward, we increased the value of the previous 30 cells. A similar approach (i.e., activity function, smoothing window, local maximum) to the construction of time series from micro-blogs (e.g., Twitter) has been followed by a growing number of researchers (e.g., see citations to Shamma et al. [7]). Finally, we construct the corresponding semantic time series (pulse-like), which models the regions of interest of a video.

In summary, the following methodology is used: 1) smoothing of aggregated the replay time-series, 2) semantic time-series, and 3) determination of time-distance between the local maximum of the user and the semantic time-series.

## RESULTS

The video segments with peaks are most likely to attract the viewers' interest. In order to determine the precise position of the peak, a derivative curve is computed. The zero-crossing points from positive to negative on derivative curve are the locations of the peaks. In this way, all key-frames in a video sequence can be identified without the need of any content-based detection. The user time series are plotted with the solid red curve and the experimentally defined ground truths are plotted with the pulse-like solid blue line.

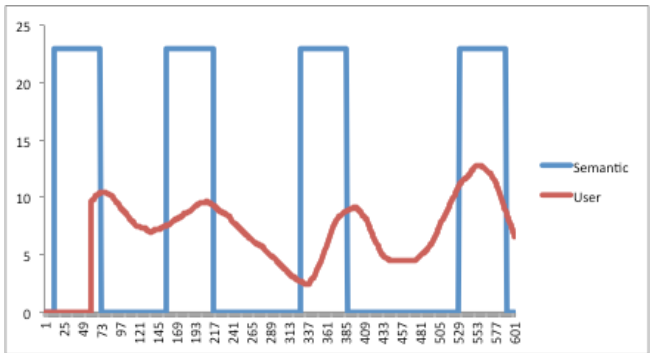

**Figure 3 The user-based interest function has accurately identified all video segments in Video A (lecture, averaging window is 60 seconds)**

We found that a simple heuristic could automatically generate video thumbnails that are positioned at the start of each interesting video segment. In order to calculate this heuristic we observed that in all cases the distance of the local maximum of the Replay30 time series from the start of the respective ground truth is less than 60 seconds. This simple heuristic detects 100% of the interesting video segments (n=8). There is only one case that the local maximum is before the start of the interesting video segment (Cooking video, S3). Therefore, we suggest that the position of user-based video thumbnails can be automatically generated for any video by locating the local maximums of the Replay30 time series and then selecting the one with the greatest value (Table 2).

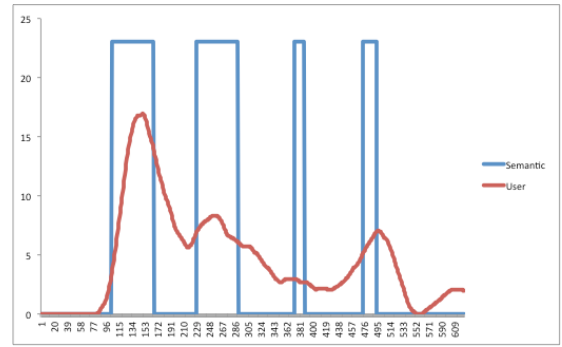

**Figure 4 The user-based interest function has been also accurate with Video B (how-to, averaging window is 45 seconds), which includes some narrow video segments**

The user interest value of a key-frame can be used as the rank of the key-frame. Based on such a measure, it is convenient to generate a ranking of importance of key-frames. Then, the maximum user interest value of the key-frames in a video could be used as its representative video thumbnail.

| Scene/Video(secs) | Lecture A | How-to B |
|---|---|---|
| S1 | 33 (40) [10] | 45 (105) **[16]** |
| S2 | 13 (145) [10] | 21 (230) [8] |
| S3 | 48 (350) [9] | -13 (374) [3] |
| S4 | 1 (554) **[13]** | 21 (475) [7] |

**Table 2 We have calculated the distance of the local Replay30 maximum from the semantic start (inside parentheses) and we provide in bold the peak values, which stand for the user-based thumbnail**

In summary, the central contribution of this work is a novel conceptualization of video data-logs that holds the following unique properties: 1) implicit from people's action, 2) video signal/content-free, 3) adaptive based on consumption. The proposed heuristic ("sixty-seconds from local maxima of user activity") explains a methodology to support our core contribution and might hold different values depending on the video and on the distribution of video interactions. Moreover, we present a reproducible method that is verified and explained via qualitative and quantitative sources. Reimplementation of this system would only require a properly instrumented video player.

## DISCUSSION AND FURTHER RESEARCH

The majority of previous approaches employed content-based (e.g., detection of object, shot, and scene change) or explicit user-based methods (e.g., comments, tags, re-mix) to improve users' watching and browsing experience. The proposed system explores the application of an implicit user-based technique. We simply record users' interactions with video player buttons. In terms of the user activity data, the most relevant work is the Audience Retention tool, which is part of the YouTube Analytics video account. The Audience Retention tool is employing the same set of data as suggested here, but there is no open documentation on the technique employed to map user interactions to a graph. Moreover, Audience Retention has been designed as a tool for video authors, but our system is proposed as a back-end tool that might improve navigation for all video viewers.

In contrast to content-based information retrieval, we have only employed four videos in the experimental procedure. Previous work on content-based information retrieval from videos has emphasized the number of videos employed in similar experiments, because the respective algorithms treated the content of those videos. In this user-based work, we are not concerned with the content of the videos, but with the user activity *within* a video. Nevertheless, it is worthwhile to explore the effect of more videos and interaction types. Therefore, the small number of videos used in the study is not an important limitation, but further research has to elaborate on different genres of video (e.g., news, sports, comedy).

Although the replay user activity seems suitable for modeling user interest, further research should consider the rest of the implicit user activities. We decided to ignore the *pause* interaction because, during the pilot tests, we noticed that the users paused the player to write down the answer to a question. Thus, the pause frequency distribution perfectly matched the ground truths, but this pattern might not have external validity. Nevertheless, in field data, a *pause* might signify an important moment, but a pause that is too long might mean that the user is away.

Another direction for further research would be to perform data mining on a large-scale web-video database. Nevertheless, we found that the experimental approach is more flexible than data mining for the development phase of the system. In particular, the iterative and experimental approach is very suitable for user-based information retrieval, because it is feasible to associate user behavior to the respective data-logs. Finally, we suggest that user-based content analysis has the benefits of continuously adapting to evolving users' preferences, as well as providing additional opportunities for the personalization of content. For example, researchers might be able to apply several personalization techniques, such as collaborative filtering, to the user activity data. In this way, implicit video pragmatics is emerging as a new playing field for improving user experience on social web multimedia.

**ACKNOWLEDGMENTS**
The author is thankful to the participants of the user study, and to David Ayman Shamma, Avlonitis Markos, Ioannis Leftheriotis, and Chryssoula Gkonela for assisting in the implementation and evaluation of the system, as well as for providing feedback on early drafts of this paper. The work reported in this paper has been partly supported by project CULT (http://cult.di.ionio.gr). CULT (MC-ERG-2008-230894) is a Marie Curie project of the European Commission (EC) under the 7th Framework Program (FP7).